\documentclass[12pt]{article}
\usepackage{latexsym}
\normalbaselines
\newcommand{\cH}{{\mathcal H}}

\begin{document}
\begin{center}
\vspace*{1.0cm}

{\large{\bf Simultaneous decompositions of two states }}

\vskip 1.5cm

Armin Uhlmann, Leipzig \footnote{postal address:
Institut f\"ur Theoretische Physik, Universit\"at Leipzig,
Augustusplatz 10, D-04109, Leipzig, Germany.
email: armin.uhlmann@itp.uni-leipzig.de}

\vskip 0.5 cm

{\bf Dedicated to Roman S. Ingarden }

\end{center}

\vspace{1 cm}

\begin{abstract}
Simultaneous decompositions
of a pair of states into pure ones are examined.
There are privileged decompositions which
are distinguished from all the other ones.
\end{abstract}

\vspace{1 cm}

Presently we witness that {\em quantum information theory}
is becoming
an inter-disciplinary, quickly growing field of research. In
its history Roman S.~Ingarden is playing a significant role,
both by his own research and by posing stimulating questions
and problems \cite{book}. It is about 40 years ago that
I met Roman the first time, and he was already thinking
about the role of information in quantum physics
and, in particular, whether one can found the concept of
probability onto that of information \cite{IU}.\\
I feel honored by the
possibility to dedicate to him the following paper.

\section{Decomposing one density operator }
A density operator, representing a state, is
a positive operator with trace one. However, it is
convenient for the following considerations not to insist
in normalization.\\
We shall assume, mainly for technical simplicity,
a finite dimensional Hilbert space, $\cH$, the
dimension  of which is denoted by $\dim \cH = d$.
Thus, mathematically,
we are just dealing with positive operators (and with the
Null operator) of a finite dimensional Hilbert space.

Let $\tau$ be a positive operator on our
Hilbert space.
 Its decreasingly ordered eigenvalues
are denoted by $\lambda_1, \lambda_2, \dots$, i.~e.
$$
\hbox{spec}(\tau) = \{ \lambda_1 \geq \lambda_2 \geq \dots \}
$$
By a {\em decomposition of $\tau$} I denote every set of
vectors $|\chi_j\rangle$ such that
\begin{equation} \label{detau1}
\tau = \sum |\chi_j \rangle\langle \chi_j|
\end{equation}
As I showed in \cite{Uh70}
\begin{equation} \label{compare1}
\sum_{j=1}^m \lambda_j \geq
\sum_{j=1}^m \langle \chi_j | \chi_j \rangle
\end{equation}
is valid for all $1 \leq m \leq \dim \cH$.
Moreover, equality is reached if and only if
$|\chi_j\rangle$ is an eigenvector for $\lambda_j$ of $\tau$
for all $j = 1, \dots, m$.\\
The motivation for asking questions of that kind has been
the problem whether the von Neumann entropy of a density
operator is already fixed by its position as a point
in the convex set of
all density operators. The result just quoted gives,
if written with normalized vectors, an
affirmative answer. Indeed, my aim was to define on every
(compact) convex set a function which just gives the
von Neumann entropy if applied to state spaces of a quantum
system. Up to day I do not know whether the construction is
of any use for other convex sets than quantum state spaces.

In \cite{Ni99}
M.~A.~Nielsen proved the reversed statement: If $p_j$ are
positive numbers which are majorized by spec$(\tau)$, then
there exists a decomposition (\ref{detau1}) such that
$p_j = \langle \chi_j | \chi_j \rangle$.

The results mentioned above will be slightly extended
to the case that there are two decompositions of one and the
same $\tau$. Thus let
\begin{equation} \label{detau2}
\tau = \sum |\chi'_j \rangle\langle \chi'_j|
\end{equation}
be a further decomposition of $\tau$. Adding (\ref{detau1})
and (\ref{detau2}) we get a decomposition of $2 \tau$ and
$$
\sum_{j=1}^m (\langle \chi_j | \chi_j \rangle  +
\langle \chi'_j | \chi'_j \rangle) \geq
2 \sum_{j=1}^m | \langle \chi_j | \chi'_j \rangle |
$$
Equality takes place iff $|\chi_j\rangle$ differs from
$|\chi_j'\rangle$ by a phase factor only. Because the
eigenvalues of $2 \tau$ are just $2 \lambda_j$
we get\\

{\bf Proposition 1: } \, {\em Let (\ref{detau1}) and (\ref{detau2})
be decompositions of
$\tau$ and $\lambda_1 \geq \lambda_2 \geq \dots$
the decreasingly ordered eigenvalues of $\tau$,
and $1 \leq m \leq d$. Then
\begin{equation} \label{detau3}
\sum_{j=1}^m \lambda_j \geq
\sum_{j=1}^m | \langle \chi_j | \chi'_j \rangle |
\end{equation}
Equality holds if and only if for $1 \leq j \leq m$
\begin{equation} \label{detau4}
\tau |\chi_j\rangle = \lambda_j |\chi_j\rangle, \quad
|\chi'_j\rangle = \epsilon_j |\chi_j\rangle
\end{equation}
with unimodular numbers $\epsilon_j$.}

\section{Decomposing two density operators }
Let us now consider a pair, $\rho$ and $\omega$, of positive
operators.\\
{\bf Definition:} \, $F^+_m(\rho, \omega)$ {\em denotes the sum of
the $m$ largest eigenvalues of}
\begin{equation} \label{F}
(\sqrt{\rho} \omega \sqrt{\rho})^{1/2}
\end{equation}
The definition works well for $1 \leq m \leq d$. It is sometimes
convenient to extend it by $F^+_m = F^+_d$ if $m \geq d$ and
to set $F^+_m = 0$ for $m=0$.\\
Remark that $F^+_d$ is the square root of the transition
probability \cite{fidel}.
The square root of the transition probability
is called {\em fidelity} and is denoted by $F(\rho, \omega)$
in the present paper. Notice, however, that
Jozsa, who showed its use in quantum information
theory \cite{Jo94}, identified the general transition
probability with his fidelity concept (and not with its
square root).\\
A further remark is the following:
In \cite{Uh00a} I considered another
quantity: The $k$-fidelity, $F_k$, which is
the sum of {\em all but the first} $k$ eigenvalues
of (\ref{F}).
These partial fidelities are jointly concave (and super-additive)
in its arguments for
$k$ = 0, 1, ...  Obviously,
$$
F_m^+(\rho, \omega) = F(\rho, \omega) - F_m(\rho, \omega)
$$
In contrast to the partial fidelities, the quantity (\ref{F+})
seems to be neither concave nor convex
if $m$ is smaller than $\dim \cH$.\\
Finally, let us rewrite (\ref{detau3}) of proposition 1 as
\begin{equation} \label{detau5}
F(\tau, \tau)  \geq
\sum_{j=1}^m | \langle \chi_j | \chi'_j \rangle |,
\end{equation}
Remember that equality in (\ref{detau5}) can be reached by
eigenvector decompositions of $\tau$ with decreasingly ordered
eigenvalues.\\

{\bf Theorem 1:} \, {\em Let be $1 \leq m \leq d$. It is
\begin{equation} \label{F+}
F^+_m(\rho, \omega) = \max \sum_{j=1}^m
|\langle \psi_j | \varphi_j \rangle|
\end{equation}
where the maximum is to perform over all possible
decompositions
\begin{equation} \label{decop}
\rho = \sum |\psi_j \rangle\langle \psi_j|, \quad
\omega = \sum |\varphi_j \rangle\langle \varphi_j| .
\end{equation}
If the length of a decomposition is less than $\dim \cH$,
or if the length of the two compositions (\ref{decop}) are
different, one adds some zero vectors to get decompositions
of equal and large enough length.}

The proof of the theorem starts by stating
the invariance of the eigenvalues of (\ref{F}) with
respect of a transformation
\begin{equation} \label{transf}
\{ \rho, \omega \} \, \Rightarrow \,
\{ \rho, \omega \}^X :=
\{ X \rho X^*, (X^{-1})^* \omega X^{-1} \}
\end{equation}
for any invertible operator $X$, see \cite{Uh00a}.
((In the present paper
the Hermitian adjoint of an operator $A$ is denoted
by $A^*$ and {\em not} by $A^{\dag}$.))
Hence the
sum of the $m$ largest eigenvalues of (\ref{F})
cannot be changed by such a transformation. On the other hand,
if we simultaneously transform decompositions (\ref{decop})
according to
\begin{equation} \label{vtrans}
|\psi_j\rangle \to X |\psi_j\rangle, \quad
|\varphi_j\rangle \to (X^{-1})^* |\varphi\rangle
\end{equation}
then the right hand side of (\ref{F+}) remains unchanged.
Therefore, if the assertion of the theorem is true for a pair
of density operators $\{ \rho, \omega \}$, it is true for
every pair $\{ \rho, \omega \}^X$.\\
Let now $\rho$ and $\omega$ be invertible (i.~e. faithful).
If we then can choose $X$ such that
\begin{equation} \label{ntransf}
\{ \rho, \omega \}^X := \{ \tau, \tau\}
\end{equation}
with a certain $\tau$ yet to be determined, we are done:
For the pair
$\{\tau, \tau\}$ the theorem is equivalent to
proposition 1. But
\begin{equation} \label{X}
X \omega X^* = (X^{-1})^* \rho  X^{-1} := \tau
\end{equation}
is valid if $X^*X$ is the geometric mean \cite{PW75}
of $\rho$ and $\omega^{-1}$, i.e.
\begin{equation} \label{geom}
X^* X =
\omega^{-1/2} \bigl( \omega^{1/2} \rho \omega^{1/2} \bigr)^{1/2}
\omega^{-1/2}
\end{equation}
Hence, the theorem is true for invertible $\rho$ and $\omega$.\\
Indeed, the proof covers the case of any pair $\rho$, $\omega$,
with equal supports: To see it we only have to replace $\cH$
by the supporting Hilbert subspace because neither to $F^+_m$
nor to the decompositions there is a non-zero contribution
from the null spaces (i.e.~the kernels) of $\rho$ and $\omega$.

We now prove that the right hand side of (\ref{F+}) never exceeds
$F^+_m$.  Denote by $P_0$, $Q_0$ the projection operators onto the null
spaces of $\rho$ and $\omega$.  We choose decompositions of $P_0$ and
$Q_0$ with vectors $|\psi'_i\rangle$ and $|\varphi'_i\rangle$
respectively.  We complement arbitrarily chosen decompositions
(\ref{decop}) to those of $\rho' = \rho + c_1 P_0$ and $\omega' = \omega
+ c_2 Q_0$ with $c_j > 0$.  For $\rho'$ this is done by $$ \rho' = \rho
+ c_1 P_0 = \sum |\psi_j \rangle\langle \psi_j| + c_1 \sum |\psi'_j
\rangle\langle \psi'_j| $$ and similarly we proceed with $\omega'$.
Because $\rho'$ and $\omega'$ are invertible, we already can apply
theorem 1 to them.  Because $F^+_m(\rho', \omega')$ is
approaching $F^+_m(\rho, \omega)$ if $c_j \to 0$ we are done.\\

What remains to show is the following: There are decompositions
(\ref{decop}) such that $\sum |\langle \psi_i, \varphi_i\rangle|$
is equal to $F^+_m$, whatsoever the support properties
of $\rho$ and $\omega$ may be. To get this we first assert:\\
Let $Q$ be the projection operator onto the supporting space
of $\omega$. For all decompositions (\ref{decop}) we get
$$
\langle Q \psi_i | \varphi\rangle =
\langle \psi_i | Q | \varphi\rangle =
\langle \psi_i | \varphi\rangle
$$
because every vector of a decomposition of $\omega$ must be
an eigenvector of $Q$. That is, every one of the sums in
question for $\rho, \omega$ gives one for $Q \rho Q, \omega$
yielding the same value. On the other hand, if we start with
decompositions  of $Q \rho Q, \omega$, we can add terms
orthogonal to $\omega$ into the decomposition of $\rho$ to
get a decompositions of $\rho, \omega$ without changing the
value of the sum. Below we shall show
the equality of $F^+_m(\rho, \omega)$ with
$F^+_m(Q \rho Q, \omega)$, and, all together, we obtain:
If and only if theorem 1 is true for the pair
$Q \rho Q, \omega$ it is true for the pair $\rho, \omega$.
Now we can proceed as following: If the supports of
$Q \rho Q$ and $\omega$ are equal, we are done. If not,
we consider the projection operator $P_1$ onto the support of
$Q \rho Q$, yielding the same statement for the pairs
$\rho, \omega$, $Q \rho Q, \omega$, and
$Q \rho Q, P_1 \omega P_1$. Either the last pair is of equal
support, and we are done, or continue the same game with the
projection operator $Q_1$ onto the support of $P_1 \omega P_1$.
This procedure must terminate after a finite number of steps
yielding a pair with equal supports. The obvious reason: In
every necessary step, the rank of one member of the pair
under consideration is diminished, and we are in finite
dimensions.\\
The proof of theorem 1 is done after showing
the equality of $F^+_m(\rho, \omega)$ with
$F^+_m(Q \rho Q, \omega)$ if $Q$ is the support projection of
$\omega$. This assertion is a particular case with $X = Q$
of the equation
\begin{equation} \label{transf3}
F^+_m(\rho, X^* \omega X) = F^+_m(X \rho X^*, \omega)
\end{equation}
For invertible $X$ the transformation (\ref{transf}) does not change the
eigenvalues of (\ref{F}).  By the replacement $\omega \to X^* \omega X$
we thus get (\ref{transf3}) for invertible $X$.  But $F^+_m$ is
continuous in its arguments, and (\ref{transf3}) is valid for
all $X$.\\

Let us underline the main point in constructing decompositions
(\ref{decop}) satisfying
\begin{equation} \label{opt}
F^+_m(\rho, \omega) = \sum_{j=1}^m \langle \psi_j | \varphi_j \rangle,
\quad
m = 1, 2, \dots
\end{equation}
We have to solve (\ref{X}) so that $X$ and $\tau$ are at our
disposal. From the spectral decomposition of $\tau$,
\begin{equation} \label{tauspec}
\tau = \sum | \chi_j \rangle\langle \chi_j |, \quad
\langle \chi_j | \chi_k \rangle = \lambda_j \delta_{jk}
\end{equation}
we get an optimal decomposition satisfying (\ref{opt}) by
\begin{equation} \label{transt}
|\psi_j\rangle = X^{-1} | \chi_j \rangle, \quad
|\varphi_j\rangle = X^* | \chi_j \rangle
\end{equation}
Such a choice fulfills the bi-orthogonal relations
\begin{equation} \label{bio}
\langle \psi_k | \varphi_j \rangle =
\langle \psi_j | \varphi_k \rangle =
\lambda_j \delta_{j k}
\end{equation}

{\bf Acknowledgement.} I like to thank P. M. Alberti, B.
Crell, Ch. Fuchs, and M. Nielsen
for valuable discussions and correspondence.



\begin{thebibliography}{CNTST}

\bibitem{book}
R.~S.~Ingarden, A.~Kossakowski, M.~Ohya, {\em Information Dynamics
and Open Systems.} Kluwer Academic Publisher 1997

\bibitem{IU}
R.~S.~Ingarden, K.~Urbanik, {\em Bull.~Acad.~Polon.~Sci.},
Ser.~Astr.~Phys., {\bf 9} 313 (1961)\\
R.~S.~Ingarden, ibidem, {\bf 11} 209 (1963)

\bibitem{Uh70}
A.~Uhlmann, {\em Rep. Math. Phys.} {\bf 1} 147 (1970)

\bibitem{Ni99}
M.~A.~Nielsen,
Probability distribution consistent with a mixed state.
quant-ph/9909020

\bibitem{Jo94}
R.~Jozsa, {\em J. Mod. Opt.} {\bf 41} 2315 (1994)


\bibitem{fidel}
A.~Uhlmann, {\em Rep. Math. Phys.} {\bf 9} 273 (1976);
P.~M.~Alberti, {\em Lett. Math. Phys.} {\bf 7} 25 (1983);
R.~Jozsa, {\em J. Mod. Opt.} {\bf 41} 2315 (1994);
Ch.~A.~Fuchs and C.~M.~Caves,
{\em Open Sys.} \& {\em Inf. Dyn.} {\bf 3} 345 (1995);
P.~M.~Alberti and A.~Uhlmann, {\em Acta Applicandae Mathematicae}
{\bf 60} 1 (2000)


\bibitem{Uh00a}
A.~Uhlmann, On partial fidelities, to appear in:
{\em Rep.~Math.~Phys.}  {\bf 45} 407 (2000)

\bibitem{PW75}
W.~Pusz and L.~Woronowicz, {\em Rep. Math. Phys.} {\bf 8}  159 (1975)


\end{thebibliography}
\end{document}